\documentclass[10pt,journal,compsoc]{IEEEtran}
\ifCLASSOPTIONcompsoc
  \usepackage[nocompress]{cite}
\else
  \usepackage{cite}
\fi
\usepackage[most]{tcolorbox}
\usepackage{tabularx}
\usepackage{url}
\usepackage{framed}
\usepackage{multirow}
\usepackage{balance}
\usepackage{booktabs}
\usepackage{fontawesome}
\usepackage{threeparttable}
\usepackage[figuresright]{rotating}
\usepackage[colorlinks=true,linkcolor=black,citecolor=black,urlcolor=black]{hyperref}
\usepackage{orcidlink}
\usepackage{caption}
\usepackage{subcaption}
\usepackage{tablefootnote}
\usepackage{float}
\usepackage{enumitem}
\usepackage{colortbl}
\usepackage{latexsym} % For the latexsym package
\usepackage{tabularx}
\usepackage{amsfonts} % For \Box command
\usepackage{amsmath}
\usepackage{array}
\usepackage{longtable}
\usepackage{supertabular}
\usepackage{xcolor}
\usepackage{soul}
\usepackage{listings}
\usepackage{seqsplit}
\usepackage{graphicx}

\definecolor{lightblue}{RGB}{220,230,241}
\definecolor{lightblue}{RGB}{173,216,230}
\definecolor{softpink}{RGB}{255,182,193}
\definecolor{mygray}{rgb}{0.9,0.9,0.9}

\hypersetup{
    colorlinks=true,
    linkcolor=blue,
    filecolor=magenta,
    urlcolor=blue,
    pdftitle={Assessing Language Models for Salient Class Identification},
    pdfpagemode=FullScreen,
}

\newcommand{\codeid}[1]{\texttt{\seqsplit{#1}}}

\begin{document}
\begin{sloppypar}

\title{Assessing Language Models for Salient Class Identification}

\author{Bo Xiong~\orcidlink{0009-0007-4559-7815}, Chaoran Cai~\orcidlink{0009-0004-7123-440X}, Zongen Ren~\orcidlink{0009-0002-9924-543X}, Kaipeng Xiong~\orcidlink{0009-0001-7890-139X}, Chong Wang~\orcidlink{0000-0003-4576-5392}, and Peng Liang~\orcidlink{0000-0002-2056-5346}
\IEEEcompsocitemizethanks{
\IEEEcompsocthanksitem Bo Xiong, Chaoran Cai, Zongen Ren, Chong Wang, and Peng Liang are with the School of Computer Science, Wuhan University, Wuhan, China.
\\ E-mail: yueshaomoon\_@whu.edu.cn, charoncai@whu.edu.cn, zongnren@whu.edu.cn, cwang@whu.edu.cn, liangp@whu.edu.cn

\IEEEcompsocthanksitem Kaipeng Xiong is with the Faculty of Information Engineering and Automation, Kunming University of Science and Technology, Kunming, China.
\\ E-mail: xkp1103@163.com}

\thanks{Manuscript received; revised.}}

\markboth{IEEE Transactions on Software Engineering}
{B. Xiong \MakeLowercase{\textit{et al}.}: Assessing Language Models for Salient Class Identification}

\IEEEtitleabstractindextext{%
\begin{abstract}

Code review requires reviewers to understand the core intent of code changes, which becomes difficult when a commit modifies multiple classes. In such commits, one or more primarily modified classes, referred to as salient classes, may induce modifications in other classes. Accurate identification of salient classes offers reviewers an effective entry point to navigate code changes and facilitates program comprehension. Existing state-of-the-art approaches rely on complex program-analysis procedures, including Abstract Syntax Tree (AST) parsing, class relation extraction, handcrafted feature engineering, or dependency graph construction. To this end, we study whether language models (LMs) can identify salient classes directly from commits without feature engineering, graph construction, or training. We first construct a new dataset ApacheJavaCM, derived from the
ApacheCM dataset, containing 7,911 commits and 25,914 labeled classes. On this dataset, we systematically evaluate whether LMs can identify salient classes directly from commits and compare with the strongest reproducible state-of-the-art (SOTA) baseline. The evaluation covers two large language models (LLMs), GPT-5.4 and DeepSeek-V3.2, one small language model (SLM), Qwen3.5-9B, and three prompting strategies: zero-shot, few-shot, and chain-of-thought. The LMs substantially outperform the baseline while remaining stable across commit characteristics and selected LMs. We also found that, for salient class identification tasks, a 9B-parameter open-source SLM, Qwen3.5-9B, under few-shot prompting, achieves performance comparable to that of a much larger closed-source LLM, GPT-5.4. These results suggest that lightweight, locally deployable SLMs are feasible options for the salient class identification task and can reduce both cost and privacy barriers associated with relying on closed-source LLMs.

\end{abstract}

}

\maketitle
\IEEEdisplaynontitleabstractindextext
\IEEEpeerreviewmaketitle

\section{Introduction}

Code review is a core software engineering practice to reduce defects and maintain software quality \cite{bacchelli2013review,thongtanunam2017participation,mcintosh2016impact,kononenko2015participation}. In both open-source communities and industrial projects, developers submit code changes to peers or maintainers for review, enabling reviewers to identify defects, improve code quality, and preserve long-term maintainability \cite{bosu2017process,kononenko2016quality,baysal2016factors,sadowski2018google}. Although code review tools such as Gerrit \cite{gerrit2026review} and CodeFlow  \cite{cacm2019codeflow} support the code review process, reviewers still spend substantial effort understanding the code changes in a commit \cite{zhang2015interactive,tao2012understand,rigby2008apache,rigby2013convergent}. This effort is further amplified when a commit spans multiple classes or files, as reviewers must reconstruct the overall intent of the change from scattered modification fragments. The challenge is particularly pronounced in large-scale or open-source projects with diverse contributor backgrounds \cite{rigby2008apache,rigby2013convergent,barnett2015decomposition,dias2015untangling,hattori2008commits}.

To address this challenge, Huang \textit{et al}.~\cite{huang2018salient,huang2022isc} proposed the concept of the \textbf{salient class}. \textcolor{black}{A commit may contain one or more salient classes: modified classes whose changes induce related modifications in other classes. Their user study showed that beginning with salient classes helps reviewers understand code changes more effectively and reduces the time required for comprehension in a reviewing scenario. Furthermore, they conducted a survey among active Gerrit reviewers and found that most respondents valued a reasonable reading order and believed that providing salient-class information as a reading cue could enhance their comprehension of code changes.} However, accurate automated identification of salient classes remains a challenge. Existing methods~\cite{huang2022isc,ren2024gbsci} are supervised learning driven by static analysis. These methods are based on program-analysis techniques such as change propagation, fine-grained source-code change extraction, and relevance identification for code changes \cite{hassan2004propagation,fluri2007change,huang2017relevance}, but they also require parsing modified classes and their surrounding class context, extracting inter-class relations, constructing code change dependency graphs over modified classes, and using the resulting representations to train a model for salient class identification.

These limitations prevent existing salient class identification methods from serving as lightweight aids for code review.
In real-world repositories, static analysis pipelines may fail when repositories contain incomplete builds, multiple programming languages, generated code, or dependencies specific to a project. In addition, retraining or adapting the salient class identification model for projects written in new programming languages incurs additional costs for data preparation and model maintenance. For both practitioners and researchers, a salient class identification method is useful only when it can be integrated into real review workflows with limited setup costs. This highlights the need for a lightweight method that identifies salient classes directly from commits without requiring repository-level static analysis or model training specific to each language.

Recent progress in pretrained language models (LMs), including large language models (LLMs) and small language models (SLMs), provides a new avenue for salient class identification task. LMs have shown strong performance in code understanding and generation tasks, such as code summarization and commit message generation \cite{feng2020codebert,dong2022fira,jiang2017nmt,dong2023revisiting,zhang2024critical}, and SLMs have become increasingly practical for local deployment and privacy-sensitive software engineering scenarios \cite{hasan2026benchmarks,zhang2024cogenesis,crupi2026slmjudge}. Unlike existing methods that rely on handcrafted features or explicit graph structures for salient class identification, LMs can directly process the code diff of a commit, infer the semantic role of each modified class through natural language prompts, and determine which classes are salient classes. In this paper, we present the first systematic empirical study of the performance of LMs in salient class identification, covering dataset construction, comparisons with SOTA reproducible baselines, analysis across LMs and prompting strategies, evaluation under commit characteristics, and qualitative analysis of identification errors.

%\textbf{Main contributions:} \textbf{(1) The first systematic evaluation of applying LMs to salient class identification.} We systematically evaluate whether LMs, including both LLMs and SLMs, can accurately identify salient classes directly from commits. \textbf{(2) A public and reproducible dataset for salient class identification.} We construct and release a reproducible dataset derived from ApacheCM, containing 13,170 commits and 47,521 labeled classes. \textbf{(3) A comprehensive experimental evaluation across LMs, prompts, and commit characteristics.} We evaluate 9 LM-prompt configurations and 3 characteristic dimensions of commits, together with statistical testing. The results show that LM-based methods outperform the current state-of-the-art reproducible baseline, and that a 9B-parameter open-source SLM under few-shot prompting can approach the performance of a closed-source LLM. \textbf{(4) A qualitative analysis of LM prediction errors.} We conduct qualitative failure analysis to attribute erroneous predictions made by LMs in salient class identification.

\textbf{Main Contributions}: (1) to provide the first systematic evaluation of the performance of LMs, including both LLMs and SLMs, to identify salient classes directly from commits; (2) to construct ApacheJavaCM, a new dataset for salient class identification derived from the ApacheCM dataset~\cite{xiong2025C3GEN}, containing 7,911 commits and 25,914 labeled classes; (3) to perform a comprehensive experimental evaluation of LM performance in salient class identification, covering nine combinations of LMs and prompting strategies under three dimensions of commit characteristics; (4) to conduct a qualitative failure analysis, investigating the underlying reasons of incorrect salient class identification by LMs.
%of LM prediction errors to investigate why LMs fail to correctly identify salient classes.

\textbf{Paper Organization}: The remainder of this paper is structured as follows: Section~\ref{sec:related} discusses the related work. Section~\ref{sec:dataset} describes the process of constructing the dataset. Section~\ref{sec:research-design} presents the experimental design. Section~\ref{sec:results} reports and analyzes the experimental results. Section~\ref{sec:threats} discusses the threats to the validity of our study. Section~\ref{sec:conclusion} concludes the paper and outlines the future directions. To facilitate reproduction and subsequent research, all data and evaluation scripts are available at~\cite{xiong2026replication}.

\section{Related Work}\label{sec:related}
This section presents the related work from three perspectives: code review and code change comprehension, salient class identification, and LMs for code change understanding.

\subsection{Code Review and Code Change Comprehension}
Code review is widely used in open-source communities and industrial projects to maintain code quality \cite{mcintosh2016impact,kononenko2015participation}. Previous studies on code review fall into two groups. Some studies have focused on tools and practices for code review. For example, Gerrit\footnote{https://www.gerritcodereview.com/} and CodeFlow\footnote{https://www.getcodeflow.com/} are two widely used tools to support developer collaboration through comment threads, review state management, static checking integration, and interaction with version control systems \cite{bosu2017process,kononenko2016quality,baysal2016factors,sadowski2018google}. Other researchers studied the cognitive burden during code review. Those empirical studies show that reviewers often need to navigate unfamiliar files, and that commits involving more files or classes usually require more time and effort to understand the intention of code changes \cite{zhang2015interactive,tao2012understand}.

Therefore, an important goal of code review assistance is to enable reviewers to quickly identify the main intent and key modifications of a commit. Existing work supports the comprehension of code changes by decomposing composite changes, clustering related modifications, identifying systematic changes, or detecting anomalous modifications \cite{barnett2015decomposition,dias2015untangling}. 
%Salient class identification has a narrower goal than these studies: it does not attempt to explain all modifications in a commit, but instead locates the class that best represents the commit intent among multiple modified classes. This class gives reviewers a more appropriate starting point for reading.

\subsection{Salient Class Identification}
% Huang \textit{et al}. first introduced the concept of salient class for code review and proposed ISC (\textbf{I}dentify \textbf{S}alient \textbf{C}lass) to locate salient classes in commits \cite{huang2018salient,huang2022isc}. They define a salient class as the primarily modified class that causes modifications in other classes within the same commit, and they show through user studies that prioritizing this class helps reviewers understand code changes more efficiently. ISC formulates salient class identification as a binary classification task at the class level. It extracts four groups of discriminative features from each commit, including structural coupling between classes, degree of code modification, commit type, and code semantic information, and then uses a random forest classifier to determine whether each modified class is salient. The main contribution of ISC is that it turns a practical code review need into a supervised learning task based on static program features.
\textcolor{black}{Huang \textit{et al}. first introduced the concept of salient class for code review and proposed ISC (\textbf{I}dentify \textbf{S}alient \textbf{C}lass)~\cite{huang2018salient,huang2022isc}. They defined salient classes as one or more prominently modified classes whose changes induce related modifications in other classes. ISC formulates this task as a binary classification problem and predicts these classes from structural coupling, code modification, commit type, and code semantic features. In the code review process, ISC supplies the identified salient classes as supplemental information, allowing reviewers to inspect them first and follow the related modifications.
Huang \textit{et al.}~\cite{huang2022isc} also found that informing reviewers about the salient classes involved in a code change can help them understand the change more quickly.}
%Huang \textit{et al.}~\cite{huang2022isc} also conducted a survey of active Gerrit reviewers and found that most respondents considered a reasonable reading order important and regarded the salient class as helpful for understanding code changes in code review practice.}

Ren et al. later proposed GBSCI (\textbf{G}raph-\textbf{B}ased \textbf{S}alient \textbf{C}lass \textbf{I}dentification) to incorporate fine-grained program dependency information into salient class identification \cite{ren2024gbsci}. GBSCI extends the definition of salient class by considering not only classes whose names appear in commit messages, but also classes that contain functions mentioned in commit messages. In its model design, GBSCI constructs dependency graphs at the sub-statement level for code before and after modification, merges them into a Code-Change Dependency Graph, and applies R-GCN to learn graph representations. It also extracts external features from class relations inspired by UML, including inheritance, implementation, and dependency relations, and combines these features with graph representations for classification. Compared with the modeling based on static features in ISC, GBSCI explicitly captures statement-level dependencies and structural relations between classes. This design based on dependencies is consistent with earlier evidence that change propagation and fine-grained change extraction are important for understanding software evolution \cite{hassan2004propagation,fluri2007change}.

Huang et al. further extended this line of work to commit message generation by introducing a generation method based on core changes \cite{huang2025corechange}. This study uses the term core class for a concept closely related to salient class, and treats core change location as a preceding step for commit message generation. It first predicts the importance of each modified class in a commit and then preferentially feeds the code changes of core classes into the generation model, thereby mitigating input truncation under long code diffs. The core class locator follows the modeling idea of ISC based on features, but revises both the feature set and the classifier. Specifically, it removes commit type features that may leak commit message information, adds code addition/deletion type features and coupling features before the change, and replaces the random forest classifier with CatBoost. Because this locator is an improved method following ISC for identifying the core class of a commit, we refer to it as ISC+ in this paper. This work shows that salient class identification can support not only code review but also downstream tasks such as commit message generation.

Taken together, ISC, GBSCI, and ISC+ establish the main baselines based on static analysis for salient class identification. However, these methods all depend on explicit feature extraction, static program analysis, graph construction, or supervised model training. In contrast, our study evaluates a different route: directly using raw code diff and candidate class lists as input and assessing whether LMs can identify salient classes without handcrafted features or program dependency graphs.

\subsection{Language Models for Code Change Understanding}
In recent years, pretrained LMs and generative LLMs have been widely used in software engineering tasks, such as code summarization and commit message generation \cite{feng2020codebert,dong2022fira,jiang2017nmt,dong2023revisiting,zhang2024critical}. These studies show that LMs can learn semantic representations from code text, natural language descriptions, and code diff sequences, and can capture the relation between code changes and development intent to some extent. Commit message generation is especially relevant because LMs must summarize the core change of a commit from code diffs.

Existing studies related to LMs mainly focus on generative tasks or general code understanding tasks, and they do not directly test whether LMs can perform salient class identification as a discriminative task at the class level. Salient class identification differs from commit message generation because it does not require a model to generate a complete natural language description. Instead, the model must judge the relative importance of multiple classes in the same commit and determine which of them are salient classes. Empirical evidence is still lacking on the effectiveness of LMs for this task, their gap from static analysis baselines, the influence of model scale and prompting strategy, and their degradation patterns under different characteristics at the commit level. Our study targets this gap.

\section{ApacheJavaCM Construction}\label{sec:dataset}
To evaluate the performance of LMs in identifying salient classes and to compare with existing SOTA reproducible baselines, we construct a new dataset, ApacheJavaCM, by curating a subset from the ApacheCM dataset~\cite{xiong2025C3GEN}. In this section, we present the dataset construction from three perspectives: the data source and labeling rules, the construction pipeline, and the demographics of ApacheJavaCM.

\subsection{Data Source and Labeling Rules}
We use the ApacheCM dataset~\cite{xiong2025C3GEN} as the initial data source. ApacheCM aggregates commit histories from numerous projects maintained by the Apache Software Foundation and contains 249,830 commits. It spans a wide range of project scales, contributor profiles, and commit styles, and has been used in several prior empirical studies on commits \cite{zhang2025codefuse, xiong2026ist}.

% During the construction of ApacheJavaCM, we apply the following labeling rules based on the definition of salient classes in \cite{huang2022isc}.
% Specifically, for a given commit, a class whose name is explicitly mentioned in the commit message is labeled as a salient class (positive sample); whereas other classes in the commit that are not mentioned in the message are labeled as non-salient classes (negative samples).

When constructing the ApacheJavaCM, we limit our data source to Java projects for three key reasons. 
%First, when labeling classes according to the above rule, the inherent CamelCase naming conventions of Java class names facilitate low-ambiguity mapping to source files and class entities, thereby improving the reliability of automatic label construction. 
First, this language scope mirrors that of previous research on salient class identification, where the datasets were mainly constructed and evaluated on Java projects. Keeping the scope of the programming language unchanged ensures that our dataset conforms to the well-established task setting in this research area. 
Second, Java projects comprise the largest language group in ApacheCM, offering both sufficient data scale and project diversity for constructing a public benchmark. Third, adopting the Java scope ensures a fair comparison with the selected ISC baseline~\cite{huang2022isc}. Since ISC depends on AST parsing for Java, semantic feature extraction, and coupling computation between classes, any data outside Java source files cannot be processed correctly.

\textcolor{black}{Following the definition of salient classes in \cite{huang2022isc}, we use two sets for each commit $c$: $D(c)$ is the set of class names extracted from the corresponding code diff in ApacheCM, and $M(c)$ is the set of Java CamelCase class names extracted from the commit message. A class is labeled as positive if its name appears in $M(c)$, and negative otherwise. Labels are assigned only to classes in $D(c)$.}

\subsection{Construction Pipeline}
Starting from 249,830 raw commits in the ApacheCM dataset, we use a filtering and labeling pipeline with eight steps to construct ApacheJavaCM. Table~\ref{tab:dataset-pipeline} summarizes each step and the corresponding change in data scale.

\begin{table*}[!htbp]
  \centering
  \caption{Dataset construction pipeline of ApacheJavaCM.}
  \label{tab:dataset-pipeline}
  \small
  \begin{tabularx}{\textwidth}{@{}r X r r@{}}
    \toprule
    \textbf{Step} & \textbf{Filtering and Labeling Rules} & \textbf{Input} & \textbf{Output} \\
    \midrule
    1 & Exclude commits from non-Java ApacheCM projects & 249,830 & 172,277 \\
    2 & Exclude commits whose messages do not mention any Java CamelCase class name & 172,277 & 45,259 \\
    %3 & Exclude commits whose messages mention all classes included by the commit, avoiding degenerate cases in which all candidate classes would be labeled as positive
    3 & \textcolor{black}{Exclude commits for which the set of class names extracted from the message is identical to the set of classes extracted from the diff ($M(c)=D(c)$)} & 45,259 & 43,109 \\
    4 & Exclude commits that involve zero or one class, retaining commits in which contrast between positive and negative samples & 43,109 & 23,856 \\
    5 & Exclude commits that do not modify any .java file, thereby removing changes limited to configuration, documentation, and metadata & 23,856 & 22,662 \\
    % 6 & Assign positive/negative labels to each class in the commit according to the definition in Section 3.1 & 22,662 & 22,662 \\
    6 & \textcolor{black}{For each class in the diff class set $D(c)$, assign a positive label if its name occurs in the message class-name set $M(c)$, and assign a negative label otherwise.} & 22,662 & 22,662 \\
    7 & Exclude commits that lack either positive or negative samples, ensuring that each remaining commit has a usable contrast between positive and negative samples & 22,662 & 13,170 \\
    8 & Exclude commits in which at least one labeled class cannot be mapped to a .java source file, ensuring compatibility with the ISC baseline & 13,170 & 7,911 \\
    \bottomrule
  \end{tabularx}
\end{table*}

% By implementing this pipeline with eight steps, ApacheJavaCM contains 7,911 commits and 25,914 labeled classes (8,382 positive and 17,532 negative samples).
\textcolor{black}{After Step 5, 22,662 commits remain. Step 6 assigns labels to the classes extracted from each diff. A class name found in the message can be absent from the diff class set, so the resulting labels can contain only negative samples. Conversely, when all modified classes are mentioned in the message, the labels contain only positive samples. Step 7 removes these one-sided label maps, leaving 13,170 commits containing both positive and negative samples. Step 8 then removes 5,259 commits whose labeled classes cannot all be mapped to Java source files, leaving the 7,911 commits in ApacheJavaCM. The released dataset contains 25,914 labeled classes: 8,382 positive and 17,532 negative.}
%with a positive-to-negative ratio of about 1 : 2.32. We will release this dataset with the replication package.

\subsection{Demographics of ApacheJavaCM}
Table~\ref{tab:dataset-demographics} summarizes the demographics of ApacheJavaCM, including the number of commits, the total number of labeled classes, the average number of classes per commit, the total number of salient classes, and the total number of non-salient classes.

\begin{table}[htbp]
  \centering
  \caption{Demographics of ApacheJavaCM.}
  \label{tab:dataset-demographics}
  \small
  \begin{tabularx}{\columnwidth}{@{}l >{\centering\arraybackslash}X@{}}
    \toprule
    \textbf{Dimension} & \textbf{ApacheJavaCM} \\
    \midrule
    Total commits & 7,911 \\
    Total labeled classes & 25,914 \\
    \hspace{1em}Total positives & 8,382 \\
    \hspace{1em}Total negatives & 17,532 \\
    Average classes per commit & 3.276 \\
    Average salient classes per commit & 1.060 \\
    Minimum / maximum classes per commit & 2 / 21 \\
    Positive-to-negative ratio & 1 : 2.09 \\
    \bottomrule
  \end{tabularx}
\end{table}

In ApacheJavaCM, most commits involve between 2 and 7 classes, with an average of 3.276 classes per commit. The dataset contains 8,382 positive samples and 17,532 negative samples, with a positive-to-negative ratio of about 1 : 2.09, indicating class imbalance in salient class identification. This distribution is consistent with the trends reported by prior works~\cite{huang2022isc,ren2024gbsci} on their reported datasets, where salient classes usually constitute only a small portion of the classes by a commit.

The average number of salient classes per commit in ApacheJavaCM is 1.060, indicating that most commits contain only one salient class. This observation is consistent with prior works~\cite{huang2022isc,ren2024gbsci}. During dataset construction, Huang \textit{et al}. \cite{huang2022isc} reported that most commits contain only one salient class. Similarly, Ren \textit{et al}. \cite{ren2024gbsci} observed during data collection that the majority of commits involve a single salient class, and further noted in their experimental analysis that this scenario is the most common case in software development. 
%Taken together, these datasets show that single-salient-class commits dominate the distribution in salient class identification.

\section{Research Design}\label{sec:research-design}
%This section introduces these research questions, the experimental dataset, the prompt design, the comparison baselines, the evaluation metrics, and the selected LMs.

\subsection{Research Questions}
%The objective of our empirical study is to assess the feasibility and effectiveness of LMs for the salient class identification task. 
\textcolor{black}{Motivated by prior findings that salient class information facilitates code change comprehension~\cite{huang2022isc}, this study empirically evaluates the feasibility and effectiveness of LMs in identifying salient classes.} Specifically, we examine whether LMs can accurately identify salient classes directly from code diffs, how their performance compares with existing SOTA reproducible baselines, and how their performance changes across different LMs, prompting strategies, and commit characteristics. To this end, we formulate four Research Questions (RQs).

\textbf{RQ1: What is the performance of LMs in the identification of salient classes?} 
This RQ is designed to assess the capability of LMs to accurately identify salient classes directly from code diffs, independent of static analysis at the repository level or supervised model training. We compare the performance of LMs with two baselines: \textbf{Most-Modified-Class}, a heuristic that identifies the class with the most modified lines as a salient class \cite{huang2022isc}, and the \textbf{ISC} framework \cite{huang2022isc}. Answering this RQ helps to examine the potential of LMs that only use code diffs in commits as a viable alternative to existing feature-based methods for identifying salient classes.

\textbf{RQ2: What is the impact of LM selection and prompting strategy on the performance of LMs in identifying the salient classes?} 
This RQ aims to investigate the impact of LM selection and prompting strategies on the performance of salient class identification by LMs. We evaluate closed-source LLMs, open-source LLMs, and open-source SLMs under zero-shot, few-shot, and chain-of-thought prompting settings. The answer of this RQ will help identify the most effective combinations of LMs and prompts for this task and determine whether lightweight and locally deployable SLMs can achieve performance comparable to that of LLMs.

\textbf{\textcolor{black}{RQ3: How do commit characteristics affect LM performance in salient class identification?}}
This RQ is designed to investigate the robustness of LM performance across three commit characteristics. \textcolor{black}{Based on the analyses conducted in ISC \cite{huang2022isc} and GBSCI \cite{ren2024gbsci}, we use (i) the total number of modified classes in a commit; (ii) the number of salient classes in a commit; and (iii) code-diff token length. The first two characteristics align with prior salient class studies, whereas the third captures the input scale presented to LMs that directly consume code diffs. The answer to RQ3 will help us understand how these commit characteristics affect LM performance in salient class identification.}
%Following ISC \cite{huang2022isc} and GBSCI \cite{ren2024gbsci}, we quantify the complexity of commits via the total number of classes and the count of salient classes in a commit, while further incorporating diff token length to account for input scale.
%The answer of RQ3 would help us understand the boundary conditions of LM performance in salient class identification and identify specific commit scenarios where performance degrades, indicating the need for additional support.

\textbf{RQ4: What factors contribute to the erroneous prediction of LMs in identifying salient classes?} 
This RQ aims to uncover the root causes behind the erroneous predictions made by LMs. We intend to perform a qualitative analysis of representative failure cases of LMs by synthesizing evidence from commit messages, code diffs, and ISC baseline outputs, \textcolor{black}{and summarize the resulting categories of root causes}. Answering this RQ can help identify the main failure patterns of LMs in salient class identification, which provide actionable insights for future improvements to methods based on LMs for salient class identification.

\begin{figure*}[!t]
  \centering
  \fontsize{6.5pt}{7.2pt}\selectfont
  \definecolor{promptheader}{RGB}{220,232,247}
  \definecolor{promptgreen}{RGB}{91,145,91}
  \definecolor{promptorange}{RGB}{224,103,38}
  \setlength{\fboxsep}{0pt}
  \newcommand{\promptbox}[2][promptgreen]{\fcolorbox{#1}{white}{\strut\hspace{0.35em}\texttt{#2}\hspace{0.35em}}}
  \newcommand{\promptblock}[2]{%
    \vspace{2pt}
    \noindent\fcolorbox{black}{white}{%
      \begin{minipage}{0.98\linewidth}
        \colorbox{promptheader}{\parbox[c][2.05em][c]{\linewidth}{\noindent\hspace{0.6em}{\small #1}}}
        \vspace{2pt}
        \hspace{0.6em}\begin{minipage}{\dimexpr\linewidth-1.2em\relax}
          #2
          \vspace{2pt}
        \end{minipage}
      \end{minipage}}
    \par}

  \promptblock{\textbf{Common System Prompt}}{%
    You are an expert software engineer. Your task is to identify the ``salient class'' from a given commit diff.\\
    % The salient class is the primary class that drives the core intent of the commit, often triggering dependent updates in other classes.\\
    \textcolor{black}{A commit may contain one or more salient classes. These are modified classes whose changes carry the central change and may be followed by related updates in other classes.}\\
    \textbf{Definition:}\\
    \texttt{-} A salient class is a Java class, interface, enum, or annotation whose changed behavior, API contract, lifecycle, state, or abstraction represents the main purpose of the commit.\\
    \texttt{-} A commit may have more than one salient class if multiple classes independently carry the core intent, or if a new abstraction and its main behavioral use are both essential to the commit.\\
    \texttt{-} Do not invent extra salient classes. If the commit has only one salient class, output only one.\\
    \textbf{Strictly filter out Ripple Effects:}\\
    \texttt{-} Callers, adapters, subclasses, or implementations that are changed only because a salient class changed its signature, return type, field type, constructor, or method contract are Ripple Effects.
  }

  \promptblock{\textbf{zero-shot prompt}}{%
    \textbf{System Prompt:} \promptbox{COMMON SYSTEM PROMPT}\\
    \textbf{Output Rules:}\\
    \texttt{-} Output only the salient class names.\\
    \texttt{-} Use simple class names only, without package names or file paths.\\
    \texttt{-} Wrap each salient class in its own \texttt{\textless core class\textgreater{} \textless/core class\textgreater} tag.\\
    \texttt{-} Do not include explanations, reasoning, markdown, bullets, or any text outside the tags.\\
    \textbf{User Prompt:}\\
    \texttt{\#\# Input Format:}\\
    \texttt{--- START OF CODE DIFF ---}\quad \promptbox{\{QUERY\_DIFF\}}\quad \texttt{--- END OF CODE DIFF ---}\\
    \texttt{\#\# Output Format:}\quad \texttt{\textless core class\textgreater ClassName\textless/core class\textgreater}
  }

  \promptblock{\textbf{few-shot prompt}}{%
    \textbf{System Prompt:} \promptbox{COMMON SYSTEM PROMPT}\\
    \textbf{Output Rules:}\\
    \texttt{-} Output only the salient class names.\\
    \texttt{-} Use simple class names only, without package names or file paths.\\
    \texttt{-} Wrap each salient class in its own \texttt{\textless core class\textgreater{} \textless/core class\textgreater} tag.\\
    \texttt{-} Do not include explanations, reasoning, markdown, bullets, or any text outside the tags.\\
    \textbf{User Prompt:}\\
    You will first see two labeled examples. Learn the distinction between salient classes and Ripple Effects from them. Then identify the salient class or classes in the target diff.\\
    \texttt{=== START OF EXAMPLE 1 ===}\quad \texttt{--- START OF CODE DIFF ---}\quad \promptbox[promptorange]{EXAMPLE 1 DIFF}\quad \texttt{--- END OF CODE DIFF ---}\\
    \texttt{--- START OF CORE CLASS OUTPUT ---}\quad \texttt{\textless core class\textgreater \promptbox[promptorange]{EXAMPLE 1 CORE CLASS}\textless/core class\textgreater}\quad \texttt{--- END OF CORE CLASS OUTPUT ---}\\
    \texttt{=== END OF EXAMPLE 1 ===}\\
    \texttt{=== START OF EXAMPLE 2 ===}\quad \texttt{--- START OF CODE DIFF ---}\quad \promptbox[promptorange]{EXAMPLE 2 DIFF}\quad \texttt{--- END OF CODE DIFF ---}\\
    \texttt{--- START OF CORE CLASS OUTPUT ---}\quad \texttt{\textless core class\textgreater  \promptbox[promptorange]{EXAMPLE 2 CORE CLASS}\textless/core class\textgreater}\quad \texttt{--- END OF CORE CLASS OUTPUT ---}\\
    \texttt{=== END OF EXAMPLE 2 ===}\\
    \texttt{=== START OF YOUR TASK ===}\quad \texttt{--- START OF CODE DIFF ---}\quad \promptbox{\{QUERY\_DIFF\}}\quad \texttt{--- END OF CODE DIFF ---}\\
    \texttt{=== END OF YOUR TASK ===}\\
    \texttt{\#\# Output Format:}\quad \texttt{\textless core class\textgreater ClassName\textless/core class\textgreater}
  }

  \promptblock{\textbf{CoT prompt}}{%
    \textbf{System Prompt:} \promptbox{COMMON SYSTEM PROMPT}\\
    \textbf{Reasoning Procedure:}\\
    Before giving the final answer, reason like a code reviewer:\\
    1. List the changed Java classes and briefly state what changed in each.\\
    2. Infer the commit's core intent from the behavioral/API/design change, not from the number of modified lines.\\
    3. Separate cause from consequence: identify which class change forced other classes to update.\\
    4. Mark test, documentation, caller-only, adapter-only, and boilerplate implementation changes as Ripple Effects.\\
    5. Select only the class or classes that carry the main intent of the commit.\\
    \textbf{Output Rules:}\\
    \texttt{-} You may write a concise reasoning section wrapped in \texttt{\textless reasoning\textgreater{} \textless/reasoning\textgreater} tags.\\
    \texttt{-} Do not use \texttt{\textless core class\textgreater} tags in the reasoning section.\\
    \texttt{-} After reasoning, write a line exactly as: \texttt{Final Answer:}\\
    \texttt{-} Under \texttt{Final Answer}, output only the salient class names wrapped in \texttt{\textless core class\textgreater{} \textless/core class\textgreater} tags.\\
    \texttt{-} Use simple class names only, without package names or file paths.\\
    \texttt{-} Wrap each salient class in its own tag.\\
    \texttt{-} Do not include any text after the final tags.\\
    \textbf{User Prompt:}\\
    \texttt{\#\# Input Format:}\\
    \texttt{--- START OF CODE DIFF ---}\quad \promptbox{\{QUERY\_DIFF\}}\quad \texttt{--- END OF CODE DIFF ---}\\
    \texttt{\#\# Output Format:}\\
    \texttt{\textless reasoning\textgreater [brief code-review reasoning following the procedure] \textless/reasoning\textgreater}\\
    \texttt{Final Answer:}\quad \texttt{\textless core class\textgreater ClassName\textless/core class\textgreater}
  }
  \caption{Construction templates for the three prompts.}
  \label{fig:prompt-design}
\end{figure*}

\subsection{Prompt Design}\label{subsec:prompt-design}
To investigate how LM selection and prompting strategy affect salient class identification in RQ2,
%we design three prompts for the experiments: zero-shot, few-shot, and chain-of-thought(CoT), as shown in Figure~\ref{fig:prompt-design}. 
we employ three prompting techniques, i.e., zero-shot, few-shot, and chain-of-thought(CoT), to design the prompts used in our experiments, as shown in Figure~\ref{fig:prompt-design}.
Each prompt asks the LM to act as an expert software engineer and identify the salient class from a given commit diff. To ensure output consistency, each predicted class must be formatted as a plain class name enclosed within a dedicated \texttt{<core class>} tag. These prompts maintain a consistent task definition, salient class definition, and Ripple Effects filtering rule. For conciseness, Figure~\ref{fig:prompt-design} consolidates these shared instructions into a common system prompt. Meanwhile, each prompt template maintains the original \texttt{System Prompt} and \texttt{User Prompt} structure by employing \texttt{[COMMON SYSTEM PROMPT]} as a placeholder. %\textcolor{black}{Because the two labeled diff examples used in the few-shot prompt are lengthy, we provide their full contents in the Appendix and represent them in Figure~\ref{fig:prompt-design} with the placeholders \texttt{WXImage DIFF} and \texttt{EntityStore DIFF}.}

\subsection{Comparison reproducible Baselines}\label{subsec:comparison-baselines}
To evaluate the ability of LMs to accurately identify salient classes directly from code diffs, without relying on static analysis at the repository level or supervised model training. Throughout this study, we compare LMs with two baselines: Most-Modified-Class and ISC \cite{huang2022isc}. Most-Modified-Class, a simple heuristic baseline used in ISC \cite{huang2022isc}, designates the class with the largest number of modified lines in a commit as the unique salient class and marks all remaining modified classes as non-salient. The second baseline is ISC \cite{huang2022isc}, which, as discussed in Section~\ref{sec:related}, is the strongest reproducible baseline. ISC is a framework based on features for identifying salient classes. It extracts multiple features from each commit, including structural coupling between classes, degree of code modification, commit type, and code semantic information, and then employs a random forest classifier to predict whether each modified class is a salient class. To reproduce ISC, we follow the reproduction protocol adopted by GBSCI~\cite{ren2024gbsci}, a prior study on salient class identification, and use the trained, directly executable ISC model released by Huang \textit{et al}. in their other work~\cite{huang2022changepatterns}.

\textcolor{black}{We also considered GBSCI~\cite{ren2024gbsci}. However, its complete replication package was unavailable at the time of our experiments, so we could not include it as a directly reproducible baseline. Accordingly, we adopted ISC~\cite{huang2022isc} as the strongest reproducible baseline in this study.}

\subsection{Evaluation Metrics}
Following prior work~\cite{huang2022isc,ren2024gbsci}, salient class identification is formulated as a binary classification problem: in each commit, every modified class must be classified as either salient or non-salient. Consistent with these prior studies, we employ \textit{Precision}, \textit{Recall}, \textit{Accuracy}, \textit{CmtAccuracy}, \textit{F1 score}, and \textit{Matthews correlation coefficient} (\textit{MCC}) as evaluation metrics, with formulas shown in  (\ref{eq:precision}) through (\ref{eq:mcc}). Salient classes are defined as positive samples and non-salient classes as negative samples. For clarity, true positive (TP), false positive (FP), true negative (TN), and false negative (FN) follow the standard binary classification definitions.

\textbf{\textit{Precision} and \textit{Recall}:} \textit{Precision} measures the proportion of correct predictions among all samples predicted as a given class, while \textit{Recall} measures the proportion of correctly identified samples among all samples that truly belong to that class. Because salient class identification involves both salient and non-salient classes, we report \textit{precision} and \textit{recall} for both classes: positive-class precision (\textit{PosPre}), negative-class precision (\textit{NegPre}), positive-class recall (\textit{PosRecall}), and negative-class recall (\textit{NegRecall}).

\textbf{\textit{Accuracy} and \textit{CmtAccuracy}:} \textit{Accuracy} measures the proportion of correctly classified class samples among all class samples and reflects overall classification performance at the class level. Since salient class labels are assigned at the commit level, we additionally adopt \textit{Commit Accuracy} (\textit{CmtAccuracy}), a metric introduced by~\cite{ren2024gbsci}, to assess whether all modified classes within a commit can be identified and classified correctly. A commit is counted as correct only if all modified classes in this commit are classified accurately. Accordingly, \textit{CmtAccuracy} measures the ratio of commits that are entirely correctly classified to the total number of commits.

\textbf{\textit{F1 score} and \textit{MCC}:} \textit{F1 score} is the harmonic mean of \textit{PosPre} and \textit{PosRecall}, balancing positive-class precision and recall. \textit{MCC} evaluates binary classification quality by considering TP, TN, FP, and FN simultaneously \cite{chicco2020mcc}. Its value ranges from $[-1, 1]$, where 1 indicates perfect prediction, 0 indicates performance comparable to random prediction, and -1 indicates completely reversed prediction.

\setcounter{equation}{0}
\begin{equation}
\label{eq:precision}
\mathrm{PosPre} = \frac{\mathrm{TP}}{\mathrm{TP} + \mathrm{FP}} \qquad \mathrm{NegPre} = \frac{\mathrm{TN}}{\mathrm{TN} + \mathrm{FN}}
\end{equation}

\begin{equation}
\label{eq:recall}
\mathrm{PosRecall} = \frac{\mathrm{TP}}{\mathrm{TP} + \mathrm{FN}} \qquad \mathrm{NegRecall} = \frac{\mathrm{TN}}{\mathrm{TN} + \mathrm{FP}}
\end{equation}

\begin{equation}
\label{eq:accuracy}
\mathrm{Accuracy} = \frac{\mathrm{TP} + \mathrm{TN}}{\mathrm{TP} + \mathrm{FP} + \mathrm{TN} + \mathrm{FN}}
\end{equation}

\begin{equation}
\label{eq:cmtaccuracy}
\mathrm{CmtAccuracy} = \frac{\#\,\mathrm{CorrectCommit}}{\#\,\mathrm{TotalCommit}}
\end{equation}

\begin{equation}
\label{eq:f1}
\mathrm{F1} = \frac{2 \times \mathrm{PosPre} \times \mathrm{PosRecall}}{\mathrm{PosPre} + \mathrm{PosRecall}}
\end{equation}

\begin{equation}
\label{eq:mcc}
\mathrm{MCC} = \frac{\mathrm{TP} \times \mathrm{TN} - \mathrm{FP} \times \mathrm{FN}}{\sqrt{(\mathrm{TP}+\mathrm{FP})(\mathrm{TP}+\mathrm{FN})(\mathrm{TN}+\mathrm{FP})(\mathrm{TN}+\mathrm{FN})}}
\end{equation}

In ApacheJavaCM, the positive-to-negative ratio is approximately 1 : 2.09, indicating a notable class imbalance. Therefore, we employ \textit{CmtAccuracy} and \textit{MCC} as the two primary metrics for answering RQ1 through RQ3. The former aligns directly with the actual consumption granularity of code reviews, while the latter assesses whether model performance remains reliable under class imbalance.

\subsection{Language Model Selection}
\label{subsec:LMs}
To evaluate the effectiveness of LM on salient class identification across different LMs, we select three representative LMs based on the following two considerations. First, both closed-source and open-source LMs should be included to account for the diverse constraints on cost, privacy, and infrastructure. Second, the selection should span a range of LM scales, allowing us to investigate whether effective identification of salient classes necessitates large models or can also be adequately performed by smaller ones. The license type, context length (Ctx Len), and LM category of each LM used in our study are summarized in Table~\ref{tab:language-models}. To ensure reproducibility and fairness, all three LMs use the same decoding configuration in all experiments, with the temperature set to 0 to minimize randomness caused by LLMs in our experiment and ensure stable and deterministic outputs.

\begin{table}[htbp]
  \centering
  \caption{LMs Used for experiments}
  \label{tab:language-models}
  \small
  \begin{tabular}{@{}l l c c@{}}
    \toprule
    \textbf{Model} & \textbf{License Type} & \textbf{Model Category} & \textbf{Ctx Len} \\
    \midrule
    GPT-5.4 & Closed-source & LLM & 1M \\
    DeepSeek-V3.2 & Open-source & LLM & 128K \\
    Qwen3.5-9B & Open-source & SLM & 262K \\
    \bottomrule
  \end{tabular}
\end{table}

\section{Results and Analysis}\label{sec:results}
This section reports the experimental results for the four RQs. All experiments use ApacheJavaCM constructed in Section~\ref{sec:dataset}. RQ1--RQ3 analyze the performance of LMs and baselines in identifying salient classes under different experimental settings, LM and prompt configurations, and commit characteristics, while RQ4 analyzes representative LM failure cases of the performance of LMs. For each RQ, we provide the experimental setting, discuss the main observations from the tables or figures, connect them to the research question, and summarize the key findings.

\begin{table*}[!htbp]
  \centering
  \caption{Comparison of LM and SOTA reproducible baseline performance for salient class identification}
  \label{tab:rq1}
  \scriptsize
  \resizebox{\textwidth}{!}{%
  \begin{tabular}{@{}l c c c c c !{\vrule width 0.4pt} c c c@{}}
    \toprule
\textbf{Method} & \textbf{PosPre} & \textbf{NegPre} & \textbf{PosRecall} & \textbf{NegRecall} & \textbf{Accuracy} & \textbf{CmtAccuracy} & \textbf{F1} & \textbf{MCC} \\
\midrule
    Most-Modified-Class & 0.3983 & 0.7149 & 0.4170 & 0.6989 & 0.6077 & 0.3610 & 0.4074 & 0.1145 \\
    ISC \cite{huang2022isc} & 0.5816 & 0.8011 & 0.5852 & 0.7988 & 0.7297 & 0.4846 & 0.5834 & 0.3833 \\
    \midrule
    GPT-5.4 (zero-shot) & \textbf{0.7644} & \textbf{0.9081} & 0.8136 & 0.8801 & \textbf{0.8586} & \textbf{0.7204} & \textbf{0.7883} & \textbf{0.6830} \\
    DeepSeek-V3.2 (zero-shot) & 0.7194 & 0.9066 & \textbf{0.8173} & 0.8476 & 0.8378 & 0.6698 & 0.7653 & 0.6452 \\
    Qwen3.5-9B (zero-shot) & 0.7572 & 0.8889 & 0.7694 & \textbf{0.8820} & 0.8456 & 0.7171 & 0.7632 & 0.6488 \\
    \bottomrule
  \end{tabular}
  }
\end{table*}

\subsection{Performance of LMs for Salient Class Identification (RQ1)}
RQ1 aims to explore whether LMs can effectively perform salient class identification using only code diffs, without static analysis at the repository level or supervised model training. For this purpose, we compare three LMs in a zero-shot setting with the two baselines, i.e., Most-Modified-Class and ISC \cite{huang2022isc}, as described in Section~\ref{subsec:comparison-baselines}. %For each LM, we select the best prompting strategy for each LM among the nine configurations in RQ2. GPT-5.4 and DeepSeek-V3.2 use zero-shot prompting, whereas Qwen3.5-9B uses few-shot prompting.
%In RQ1, we evaluate all three LMs under the zero-shot prompting setting.

\subsubsection{Results of RQ1}
Table~\ref{tab:rq1} reports the performance of the three LMs and the two baselines in identifying salient classes on ApacheJavaCM. Among the two baselines, ISC substantially outperforms Most-Modified-Class, achieving 0.4846 \textit{CmtAccuracy} and 0.3833 \textit{MCC}, compared to 0.3610 and 0.1145 for the Most-Modified-Class approch.

All three LMs consistently outperform both baselines in salient class identification. More specifically, GPT-5.4 achieves the highest \textit{CmtAccuracy}, \textit{MCC} and \textit{F1}, while DeepSeek-V3.2 achieves the highest \textit{PosRecall}, and Qwen3.5-9B  achieves the highest \textit{NegRecall}. Although DeepSeek-V3.2 underperforms relative to the other two LMs, it still substantially exceeds both Most-Modified-Class and ISC.

\subsubsection{Analysis of RQ1}

\textbf{LMs outperform existing SOTA baselines in salient class identification, without relying on feature engineering or graph construction.} All three LMs outperform both Most-Modified-Class and ISC across all eight metrics. Even DeepSeek-V3.2, the weakest performer among the selected LMs on the two primary metrics, improves upon ISC by 18.52\% in \textit{CmtAccuracy} and 26.19\% in \textit{MCC}. Compared to Most-Modified-Class, the corresponding improvements are 30.88\% and 53.07\%, respectively. Unlike ISC requires AST parsing, coupling extraction between classes, and random forest training, the LMs take only the raw code diff of each commit as input. These findings indicate that LMs not only achieve higher identification quality in salient class identification but also reduce the cost of static analysis at the repository level.

\textbf{Open-source LMs perform nearly as well as closed-source LMs on salient class identification.} As a closed-source LLM, GPT-5.4 achieves the highest \textit{MCC}, \textit{PosPre}, and \textit{F1}, reflecting its strongest overall classification quality. The two open-source LMs remains highly competitive: DeepSeek-V3.2 reaches 0.6698 \textit{CmtAccuracy} and 0.6452 \textit{MCC}, while Qwen3.5-9B attains 0.7171 \textit{CmtAccuracy} and 0.6488 \textit{MCC}. The comparison results suggests that, for salient class identification, open-source LMs exhibit performance approaching that of closed-source LMs.

\textbf{On salient class identification, SLMs achieve performance comparable to that of LLMs.} As a 9B-parameter open-source SLM, Qwen3.5-9B performs competitively against GPT-5.4, with only a 0.33\% gap in \textit{CmtAccuracy} and a 3.42\% gap in \textit{MCC}, while slightly exceeding GPT-5.4 on \textit{NegRecall}. Qwen3.5-9B also outperforms DeepSeek-V3.2 in \textit{CmtAccuracy}, \textit{NegRecall}, \textit{Accuracy}, and \textit{MCC}. \textcolor{black}{Since Qwen3.5-9B can be deployed locally while achieving comparable benchmark performance, it is a feasible option for this task, especially in settings when lower cost or stronger code privacy protection is desired.}
%Since Qwen3.5-9B can be deployed locally, lightweight SLMs are already a practical option for this relatively focused code understanding task, maintaining high prediction quality while reducing both cost and code privacy risk.

% \noindent\begin{center}
%   \begin{tcolorbox}[colback=black!5, colframe=black!20, width=1.0\linewidth, arc=0.5mm, boxrule=0.8pt, left=1mm, right=1mm, top=0.5mm, bottom=0.5mm, boxsep=0.5mm]
%     \textbf{Key Findings of RQ1:} LMs consistently outperform existing baseline methods for salient class identification. Notably, even the weakest LM still surpasses the best reproducible baseline, ISC. Open-source LLMs achieve performance comparable to closed-source LLMs on salient class identification. As an open-source SLM, Qwen3.5-9B performs competitively with the LLM GPT-5.4 and outperforms the LLM DeepSeek-V3.2 in \textit{CmtAccuracy}. This suggests that locally deployable SLMs are a practical option for salient class identification.
%   \end{tcolorbox}
% \end{center}
\noindent\begin{center}
  \begin{tcolorbox}[colback=black!5, colframe=black!20, width=1.0\linewidth, arc=0.5mm, boxrule=0.8pt, left=1mm, right=1mm, top=0.5mm, bottom=0.5mm, boxsep=0.5mm]
    \textbf{Key Findings of RQ1:}
    \begin{itemize}[leftmargin=*, nosep]
      \item LMs consistently outperform existing baseline methods for salient class identification. Notably, even the weakest LM still surpasses the best reproducible baseline, ISC.
      \item Open-source LLMs achieve performance comparable to closed-source LLMs on salient class identification.
      \item As an open-source SLM, Qwen3.5-9B performs competitively with the LLM GPT-5.4 and outperforms the LLM DeepSeek-V3.2 in \textit{CmtAccuracy.} This suggests that locally deployable SLMs are a practical option for salient class identification.    \end{itemize}
  \end{tcolorbox}
\end{center}

\begin{table*}[!htbp]
  \centering
  \caption{Performance of three LMs under three prompting strategies}
  \label{tab:rq2}
  \scriptsize
  \resizebox{\textwidth}{!}{%
  \begin{tabular}{@{}l l c c c c c !{\vrule width 0.4pt} c c c@{}}
    \toprule
    \textbf{LM} & \textbf{Prompt} & \textbf{PosPre} & \textbf{NegPre} & \textbf{PosRecall} & \textbf{NegRecall} & \textbf{Accuracy} & \textbf{CmtAccuracy} & \textbf{F1} & \textbf{MCC} \\
    \midrule
    \multirow{3}{*}{GPT-5.4}
      & zero-shot & 0.7644 & 0.9081 & 0.8136 & 0.8801 & \textbf{0.8586} & 0.7204 & \textbf{0.7883} & \textbf{0.6830} \\
      & few-shot  & 0.7525 & \textbf{0.9107} & \textbf{0.8213} & 0.8709 & 0.8548 & 0.7124 & 0.7854 & 0.6775 \\
      & CoT       & 0.7593 & 0.9084 & 0.8151 & 0.8765 & 0.8566 & 0.7198 & 0.7862 & 0.6795 \\
    \midrule
    \multirow{3}{*}{DeepSeek-V3.2}
      & zero-shot & 0.7194 & 0.9066 & 0.8173 & 0.8476 & 0.8378 & 0.6698 & 0.7653 & 0.6452 \\
      & few-shot  & 0.7317 & 0.8987 & 0.7972 & 0.8603 & 0.8399 & 0.6897 & 0.7630 & 0.6438 \\
      & CoT       & 0.7362 & 0.8934 & 0.7839 & 0.8657 & 0.8393 & 0.6911 & 0.7593 & 0.6396 \\
    \midrule
    \multirow{3}{*}{Qwen3.5-9B}
      & zero-shot & 0.7572 & 0.8889 & 0.7694 & 0.8820 & 0.8456 & 0.7171 & 0.7632 & 0.6488 \\
      & few-shot  & \textbf{0.7668} & 0.8860 & 0.7607 & \textbf{0.8894} & 0.8478 & \textbf{0.7232} & 0.7637 & 0.6515 \\
      & CoT       & 0.7547 & 0.8883 & 0.7683 & 0.8806 & 0.8443 & 0.7102 & 0.7615 & 0.6460 \\
    \bottomrule
  \end{tabular}
  }
\end{table*}

\subsection{Impact of LM Types and Prompting Strategies (RQ2)}
To answer RQ2, this subsection examines which type of LM, when paired with which prompting technique, performs best for the automatic identification of salient classes. In this paper, we evaluate a total of nine configurations that arise from combining three LMs (see Section~\ref{subsec:LMs}) with three prompting strategies (see Section~\ref{subsec:prompt-design}). More specifically, the zero-shot prompt consists solely of the task description and the definition of the salient class. 
%Few-shot adds two manually selected representative diff-to-salient-class examples to the zero-shot. 
Few-shot adds two representative diff-to-salient-class examples selected following the example selection process used in the prior work~\cite{feng2024adbgpt}. The selection process involved three human evaluators, including two PhD students in software engineering and one industrial software engineering expert, who independently nominated suitable examples and then discussed, merged, and refined the candidates until reaching an agreement on the final two examples. %CoT requires the LM to derive its answer via a reasoning chain with five steps.
CoT prompting further asks the LM to identify salient classes through an explicit code-review-style reasoning process before outputting its final predictions. As shown in Figure~\ref{fig:prompt-design}, the reasoning procedure consists of five steps: listing the changed Java classes, summarizing the modification in each class, inferring the core intent of the commit, separating root-cause changes from downstream Ripple Effects, and finally selecting only the class or classes that carry the main intent. This design encourages the LM to compare candidate classes by their semantic roles rather than by superficial signals such as the number of modified lines.
%RQ2 compares different model and prompting strategy combinations, addressing two practical questions: which type of model should be selected, and which prompting strategy is more suitable for salient class identification. We evaluate 9 configurations formed by 3 LMs, GPT-5.4 as a closed-source LLM, DeepSeek-V3.2 as an open-source LLM, and Qwen3.5-9B as an open-source SLM, together with the three prompting strategies introduced in Section \ref{subsec:prompt-design}: zero-shot, few-shot, and chain-of-thought(CoT). Zero-shot provides only the task description and the definition of the salient class. Few-shot adds 2 manually selected representative diff-to-salient-class examples to zero-shot. CoT requires the model to produce an answer through the steps of class summary, change analysis, dependency reasoning, and conclusion. 

\subsubsection{Results of RQ2}
Table~\ref{tab:rq2} reports a comprehensive summary of LM performance in salient class identification across nine configurations (3 LMs $\times$ 3 prompting strategies). Across all these nine configurations, \textit{Accuracy} consistently exceeds 0.8300, while \textit{CmtAccuracy} and \textit{MCC} reach at least 0.6698 and 0.6396, respectively. Compared with ISC performance reported in Table~\ref{tab:rq1}, even the weakest LM configuration surpasses the ISC baseline (Table~\ref{tab:rq1}) by 18.52\% in CmtAccuracy and 25.63\% in MCC.

Regarding LM variants, GPT-5.4 with zero-shot prompting achieves the highest \textit{Accuracy}, \textit{F1}, and \textit{MCC}, reaching 0.6830 in \textit{MCC}. Qwen3.5-9B with few-shot prompting achieves the highest \textit{CmtAccuracy}, \textit{PosPre}, and \textit{NegRecall}, with \textit{CmtAccuracy} peaking at 0.7232. DeepSeek-V3.2 achieves its highest \textit{CmtAccuracy} under CoT (0.6911) and its highest \textit{MCC} under zero-shot (0.6452).

From the perspective of prompting strategies, GPT-5.4 exhibits \textit{CmtAccuracy} ranging from 0.7124 to 0.7204 and \textit{MCC} from 0.6775 to 0.6830. DeepSeek-V3.2 shows \textit{CmtAccuracy} between 0.6698 and 0.6911 and \textit{MCC} between 0.6396 and 0.6452. Qwen3.5-9B displays \textit{CmtAccuracy} from 0.7102 to 0.7232 and \textit{MCC} from 0.6460 to 0.6515. Notably, for Qwen3.5-9B, few-shot achieves both its highest \textit{CmtAccuracy} (0.7232) and \textit{MCC} (0.6515), whereas CoT yields the lowest, i.e., 0.7102 in \textit{CmtAccuracy} and 0.6460 in \textit{MCC}.

\subsubsection{Analysis of RQ2}
\textbf{The performance of LLMs on salient class identification remains relatively stable across prompting strategies, whereas the SLM depends more heavily on guidance from few-shot examples.} For the two LLMs, i.e., GPT-5.4 and DeepSeek-V3.2, their performance on the salient class identification task varies only slightly across the three prompting strategies. GPT-5.4 has an \textit{MCC} variation of only 0.55\%, from 0.6775 to 0.6830, and a \textit{CmtAccuracy} variation of only 0.80\%, from 0.7124 to 0.7204. DeepSeek-V3.2 similarly has an \textit{MCC} variation of only 0.56\%, from 0.6396 to 0.6452. In contrast, Qwen3.5-9B, the SLM, achieves its best performance under few-shot prompting, with the highest \textit{CmtAccuracy} and \textit{MCC}. This result indicates that LLMs can directly identify the salient classes from a commit based on the task definition, while Qwen3.5-9B benefits more from concrete few-shot examples. Such examples help the SLM align with the task of which classes should be regarded as the salient classes.

\textbf{CoT does not bring stable benefits to LMs for salient class identification.} CoT achieves the highest \textit{CmtAccuracy} for DeepSeek-V3.2, 0.6911, but its \textit{MCC} is lower than that of zero-shot prompting. For GPT-5.4, CoT performs almost the same as zero-shot prompting. For Qwen3.5-9B, CoT is lower than few-shot prompting. These results indicate that LMs do not consistently benefit from explicit long reasoning chains in salient class identification. Thus, the task requires the LM to locate the salient classes in the commit rather than on constructing lengthy intermediate reasoning, which can introduce additional decision points without improving the final performance.

\textbf{For salient class identification, open-source SLMs already have practical application potential.} Table~\ref{tab:rq2} shows that, under the best prompt for each LM, Qwen3.5-9B achieves the highest \textit{CmtAccuracy}, 0.7232, followed by GPT-5.4, 0.7204, and DeepSeek-V3.2, 0.6911. For \textit{MCC}, GPT-5.4 achieves the highest value, 0.6830, followed by Qwen3.5-9B, 0.6515, and DeepSeek-V3.2, 0.6452. Although the two metrics rank the LMs differently, both results support the same conclusion: Qwen3.5-9B, an SLM, achieves performance comparable to that of GPT-5.4, an LLM, on salient class identification. Since SLM can be deployed locally, this result suggests that open-source SLMs with lower cost can serve as practical alternatives to LLMs for salient class identification, especially in scenarios that require lower cost or stronger code privacy protection.

\noindent\begin{center}
  \begin{tcolorbox}[colback=black!5, colframe=black!20, width=1.0\linewidth, arc=0.5mm, boxrule=0.8pt, left=1mm, right=1mm, top=0.5mm, bottom=0.5mm, boxsep=0.5mm]
    \textbf{Key Findings of RQ2:}
    \begin{itemize}[leftmargin=*, nosep]
    \item In the salient class identification task, LLMs show only small performance differences across the three prompting strategies; in contrast, the SLM benefits more clearly from few-shot examples. Moreover, CoT does not improve the performance of LMs on this task.
    \item For salient class identification, the SLM achieves performance comparable to that of LLMs, suggesting that it can serve as a practical alternative for this task.
    \end{itemize}
  \end{tcolorbox}
\end{center}

\subsection{Salient Class Identification Performance across Commit Characteristics (RQ3)}
RQ3 analyzes the impact of characteristics of commits on salient class identification performance. Following prior work~\cite{huang2022isc,ren2024gbsci}, we first examine the effects of the total number of classes and the number of salient classes in a commit. Because LMs directly read code diff, we also analyze the effect of code diff token length on salient class identification performance. We therefore perform grouped evaluation along three dimensions: the total number of classes in a commit, the number of salient classes, and the code diff token length. First, based on the number of classes in a commit, we divide commits into three groups: c1, commits with 2--4 classes; c2, commits with 5--10 classes; and c3, commits with at least 11 classes. Second, based on the number of salient classes in a commit, we divide commits into three groups: sc1, commits with exactly one salient class; sc2, commits with exactly two salient classes; and sc3, commits with at least three salient classes. Third, to investigate how LM performance in salient class identification changes across commits with different lengths, we measure the length of each commit by the number of tokens in its code diff. We sort all commits in ApacheJavaCM by diff token length and split the ordered commits into three parts of approximately equal size. The two resulting split points are 1,045 and 1,823 tokens: roughly one third of the commits have diff lengths no greater than 1,045 tokens, and roughly two thirds have diff lengths no greater than 1,823 tokens. For readability and interpretability, we round these values to 1,000 and 2,000 tokens and use them as the group boundaries. These cut points allow us to evaluate how LM performance in salient class identification varies across different diff token lengths. Accordingly, we divide commits into three groups: dl1, commits with no more than 1,000 tokens; dl2, commits with 1,001--2,000 tokens; and dl3, commits with more than 2,000 tokens. 
%Third, for diff token length, we use the \texttt{cl100k\_base} tokenizer in \texttt{tiktoken} as a deterministic proxy for input length that is independent of LM choice. We follow two thresholds from Google engineering practices \cite{google2026smallcls}: no more than 100 lines of code (LOC), approximately 1,000 tokens, is considered reasonable, and no more than 200 LOC, approximately 2,000 tokens, is considered a small changelist (CL). We therefore divide commits into three groups: dl1, commits with no more than 1,000 tokens; dl2, cssommits with 1,001--2,000 tokens; and dl3, commits with more than 2,000 tokens.

\subsubsection{Results of RQ3}
Tables~\ref{tab:rq3-classes}--\ref{tab:rq3-diff-length} present the salient class identification performance of the LMs and baselines on commits grouped by the three characteristics of commits defined above. For each group, the table reports \textit{Acc}, \textit{CmtAccuracy}, and \textit{MCC} as separate subcolumns. In all groups, the three LM configurations have higher \textit{CmtAccuracy} and \textit{MCC} than Most-Modified-Class and ISC.

Table~\ref{tab:rq3-classes} reports the performance of LMs and baselines for salient class identification, grouped by the total number of classes in a commit. LMs have higher \textit{CmtAccuracy} in the c1 (2--4 classes) and c3 (at least 11 classes) groups than in the c2 (5--10 classes) group. For example, the \textit{CmtAccuracy} values of GPT-5.4 zero-shot on c1, c2, and c3 are 0.7425, 0.6088, and 0.7263, respectively, while those of Qwen3.5-9B few-shot are 0.7446, 0.6165, and 0.7053. At the same time, the \textit{Accuracy} of all methods increases as the total number of classes in a commit increases. For example, GPT-5.4 zero-shot rises from 0.8350 to 0.8902 and then to 0.9683.

\begin{table*}[!t]
  \centering
  \caption{Performance of three LMs for salient class identification under three prompting strategies}
  \label{tab:rq3-classes}
  \scriptsize
  \resizebox{\textwidth}{!}{%
  \begin{tabular}{@{}l c c c c c c c c c@{}}
    \toprule
\multirow{2}{*}{\textbf{Method}} & \multicolumn{3}{c}{\textbf{c1}} & \multicolumn{3}{c}{\textbf{c2}} & \multicolumn{3}{c}{\textbf{c3}} \\
\cmidrule(lr){2-4}\cmidrule(lr){5-7}\cmidrule(l){8-10}
& \textbf{Acc} & \textbf{CmtAccuracy} & \textbf{MCC} & \textbf{Acc} & \textbf{CmtAccuracy} & \textbf{MCC} & \textbf{Acc} & \textbf{CmtAccuracy} & \textbf{MCC} \\
\midrule
    Most-Modified-Class & 0.5220 & 0.3733 & 0.015 & 0.7552 & 0.3009 & 0.204 & 0.7964 & 0.3368 & 0.208 \\
    ISC & 0.6759 & 0.5051 & 0.331 & 0.8099 & 0.4012 & 0.360 & 0.9263 & 0.4632 & 0.528 \\
    GPT-5.4 (zero-shot) & \textbf{0.8350} & 0.7425 & \textbf{0.662} & \textbf{0.8902} & 0.6088 & \textbf{0.652} & \textbf{0.9683} & \textbf{0.7263} & \textbf{0.809} \\
    DeepSeek-V3.2 (zero-shot) & 0.8123 & 0.6931 & 0.622 & 0.8714 & 0.5494 & 0.603 & 0.9596 & 0.7158 & 0.775 \\
    Qwen3.5-9B (few-shot) & 0.8202 & \textbf{0.7446} & 0.628 & 0.8870 & \textbf{0.6165} & 0.616 & 0.9604 & 0.7053 & 0.750 \\
    \bottomrule
  \end{tabular}
  }
\end{table*}

Table~\ref{tab:rq3-salient-count} reports the performance of LMs and baselines for salient class identification across groups defined by the number of salient classes in a commit. From sc1 to sc2 / sc3, the \textit{CmtAccuracy} of Most-Modified-Class, ISC, and the three LM configurations all decrease. For example, GPT-5.4 zero-shot drops from 0.7439 on sc1 to 0.3260 on sc2 and 0.2333 on sc3. DeepSeek-V3.2 zero-shot drops from 0.6902 to 0.3309 and 0.2000. Qwen3.5-9B few-shot drops from 0.7532 to 0.2157 and 0.1333.

\begin{table*}[!t]
  \centering
  \caption{Performance over Different Salient Class Counts}
  \label{tab:rq3-salient-count}
  \scriptsize
  \resizebox{\textwidth}{!}{%
  \begin{tabular}{@{}l c c c c c c c c c@{}}
    \toprule
\multirow{2}{*}{\textbf{Method}} & \multicolumn{3}{c}{\textbf{sc1}} & \multicolumn{3}{c}{\textbf{sc2}} & \multicolumn{3}{c}{\textbf{sc3}} \\
\cmidrule(lr){2-4}\cmidrule(lr){5-7}\cmidrule(l){8-10}
& \textbf{Acc} & \textbf{CmtAccuracy} & \textbf{MCC} & \textbf{Acc} & \textbf{CmtAccuracy} & \textbf{MCC} & \textbf{Acc} & \textbf{CmtAccuracy} & \textbf{MCC} \\
\midrule
    Most-Modified-Class & 0.6088 & 0.3798 & 0.117 & 0.6055 & 0.0417 & 0.158 & 0.4833 & 0.0333 & -0.016 \\
    ISC & 0.7359 & 0.5060 & 0.397 & 0.6746 & 0.1250 & 0.320 & 0.5000 & 0.0667 & 0.026 \\
    GPT-5.4 (zero-shot) & \textbf{0.8651} & 0.7439 & \textbf{0.697} & \textbf{0.7890} & 0.3260 & \textbf{0.568} & \textbf{0.7444} & \textbf{0.2333} & \textbf{0.519} \\
    DeepSeek-V3.2 (zero-shot) & 0.8438 & 0.6902 & 0.660 & 0.7781 & \textbf{0.3309} & 0.543 & 0.6889 & 0.2000 & 0.411 \\
    Qwen3.5-9B (few-shot) & 0.8560 & \textbf{0.7532} & 0.669 & 0.7640 & 0.2157 & 0.522 & 0.6500 & 0.1333 & 0.357 \\
    \bottomrule
  \end{tabular}
  }
\end{table*}

Table~\ref{tab:rq3-diff-length} reports the performance of LMs and baselines for salient class identification across groups defined by diff token length in a commit. For the three LM configurations, \textit{CmtAccuracy} decreases as diff token length increases. GPT-5.4 zero-shot drops from 0.7698 on dl1 to 0.7362 on dl2 and then to 0.6418 on dl3. DeepSeek-V3.2 zero-shot drops from 0.7290 to 0.6777 and then to 0.5922. Qwen3.5-9B few-shot drops from 0.7698 to 0.7353 and then to 0.6532. In contrast, the \textit{Accuracy} of the three LM configurations increases as the code diff token length increases.

\begin{table*}[!t]
  \centering
  \caption{performance over different diff token length}
  \label{tab:rq3-diff-length}
  \scriptsize
  \resizebox{\textwidth}{!}{%
  \begin{tabular}{@{}l c c c c c c c c c@{}}
    \toprule
\multirow{2}{*}{\textbf{Method}} & \multicolumn{3}{c}{\textbf{dl1}} & \multicolumn{3}{c}{\textbf{dl2}} & \multicolumn{3}{c}{\textbf{dl3}} \\
\cmidrule(lr){2-4}\cmidrule(lr){5-7}\cmidrule(l){8-10}
& \textbf{Acc} & \textbf{CmtAccuracy} & \textbf{MCC} & \textbf{Acc} & \textbf{CmtAccuracy} & \textbf{MCC} & \textbf{Acc} & \textbf{CmtAccuracy} & \textbf{MCC} \\
\midrule
    Most-Modified-Class & 0.4699 & 0.3845 & -0.059 & 0.5718 & 0.3581 & 0.066 & 0.7095 & 0.3391 & 0.182 \\
    ISC & 0.6576 & 0.5224 & 0.331 & 0.7083 & 0.4902 & 0.364 & 0.7853 & 0.4342 & 0.388 \\
    GPT-5.4 (zero-shot) & \textbf{0.8280} & \textbf{0.7698} & \textbf{0.656} & \textbf{0.8526} & \textbf{0.7362} & \textbf{0.682} & \textbf{0.8795} & 0.6418 & \textbf{0.667} \\
    DeepSeek-V3.2 (zero-shot) & 0.8086 & 0.7290 & 0.622 & 0.8288 & 0.6777 & 0.639 & 0.8607 & 0.5922 & 0.628 \\
    Qwen3.5-9B (few-shot) & 0.8116 & \textbf{0.7698} & 0.622 & 0.8387 & 0.7353 & 0.646 & 0.8741 & \textbf{0.6532} & 0.637 \\
    \bottomrule
  \end{tabular}
  }
\end{table*}

\subsubsection{Analysis of RQ3}

\textbf{LM performance on salient class identification does not decline monotonically as the total number of classes within a commit increases.} 
Across c1 (2--4 classes), c2 (5--10 classes) and c3 (at least 11 classes), the \textit{CmtAccuracy} of the three LM configurations follows a U-shaped pattern, and ISC shows the same tendency. For GPT-5.4, \textit{CmtAccuracy} is 0.7425, 0.6088, and 0.7263 in c1, c2, and c3, respectively. \textit{MCC} also exhibits a non-monotonic trend, first decreasing from 0.662 in c1 to 0.652 in c2, then rising to 0.809 in c3 for GPT-5.4. We therefore observe that medium sized commits are the most challenging group for LMs in the salient class identification task. A possible reason could be structural ambiguity. That is, small commits contain few candidate classes, while large commits often exhibit a clearer dominant change that propagates to many related classes. Medium-size commits, in contrast, may include several classes with comparable roles, making the salient classes less obvious.

\textbf{Although \textit{Accuracy} increases with both the total number of classes in a commit and the diff token length, this does not imply that salient class identification becomes progressively easier for LMs.} The monotonic increase in \textit{Accuracy} is better attributed to an imbalance effect. From c1 (2--4 classes) to c3 (at least 11 classes), the \textit{Accuracy} of GPT-5.4 rises from 0.8350 to 0.8902 and then to 0.9683; a similar upward trend is observed as diff token length grows. In the salient class identification, salient classes are usually fewer than non-salient classes within a commit~\cite{ren2024gbsci}. When a commit contains more classes or a longer code diff, the additional candidate classes are more likely to be non-salient classes, thereby increasing the proportion of negative samples. As a result, negative-sample decisions can dominate the overall \textit{Accuracy} calculation and raise the value of this metric, while positive-class recognition still remains difficult. 
For this reason, \textit{Accuracy} should be considered as a secondary metric when evaluating the performance of LMs on salient class identification across commit characteristics (RQ3), whereas \textit{CmtAccuracy} and \textit{MCC} serve as the primary evidence of salient class identifiation ability.

%Unlike the U-shaped or decreasing trend of CmtAccuracy, the Accuracy of all methods increases monotonically from c1 to c2 and c3. For example, GPT-5.4 zero-shot rises from 0.8350 to 0.8902 and then to 0.9683. A similar pattern appears along the dimension of diff token length. The increase mainly reflects stronger class imbalance as input scale grows. When a commit involves more classes or a longer code diff, the number of non-salient classes usually increases, while the number of salient classes remains limited. As long as a method maintains high recognition precision on negative samples, overall Accuracy naturally increases, while the true difficulty of positive sample recognition is masked. This observation is consistent with the reports of ISC \cite{huang2022isc} and GBSCI \cite{ren2024gbsci}, and it further shows that Accuracy alone cannot fully reflect salient class identification ability. CmtAccuracy and MCC are more appropriate primary metrics.

\textbf{LM performance on salient class identification declines as the number of salient classes in a commit increases.} From sc1 (one salient class in a commit) to sc2 (two salient classes in a commit) and sc3 (at least three salient classes in a commit), the three LM configurations all decline in \textit{Accuracy}, \textit{CmtAccuracy}, and \textit{MCC}. The two baselines show the same overall degradation. These results indicate that salient class identification becomes more difficult when a commit contains more salient classes. For each commit, missing even one salient class renders the entire commit incorrect; therefore, commits with multiple salient classes impose a much stricter requirement on model identification performance than those with a single salient class. This strict criterion for each commit explains why \textit{CmtAccuracy} drops more sharply than \textit{Accuracy} and \textit{MCC}. The degradation is especially clear for Qwen3.5-9B, indicating that the SLM performs worse than the two LLMs when multiple salient classes need to be identified in the same commit.
%The impact is strongest for the SLM. Qwen3.5-9B shows the largest decline among the three LMs across all three reported metrics.

%Along the dimension of the number of salient classes, all methods degrade clearly on sc2 and sc3. From sc1 to sc2 / sc3, the CmtAccuracy of Most-Modified-Class, ISC, and the three LM configurations decreases markedly. This degradation trend is consistent with the trend observed by GBSCI on its original dataset, showing that commits with multiple salient classes are an inherent challenge of the task rather than a weakness of a specific method. Any method that adopts independent binary classification for each modified class will find it harder to correctly identify all classes when multiple parallel salient classes exist in the same commit. Qwen3.5-9B, as an SLM, degrades more strongly on commits with multiple salient classes. This is consistent with its behavior of predicting only 1.060 classes per commit on average, the most conservative among the three LMs. Its tendency to output only one salient class tends to reduce PosRecall on commits with multiple salient classes.

\textbf{A longer diff token length tends to degrade the performance of LMs in salient class identification.}
As diff token length increases, LM performance on salient class identification declines: \textit{CmtAccuracy} decreases monotonically for all LM configurations, dropping by 11.66\% to 13.68\% from dl1 (no more than 1,000 tokens) to dl3 (more than 2,000 tokens). When a code diff becomes longer, the few fragments that carry the core modification semantics are surrounded by more formatting changes, interface updates, test synchronizations, or peripheral edits. The LM must therefore identify which modified classes are salient class from a larger candidate class set. Although \textit{Accuracy} also increases as diff token length grows, this increase does not indicate that salient class identification becomes easier. Longer code diffs usually contain more modified classes, and most of these additional classes are non-salient. Since non-salient classes are easier to classify correctly than salient classes, the growing number of non-salient classes contributes more correct classifications at the class level to \textit{Accuracy}. As a result, \textit{Accuracy} may increase even when identifying all salient classes in a commit remains challenging. Therefore, the effect of diff token length is better reflected by \textit{CmtAccuracy}, because this metric counts a commit as correct only when all modified classes in that commit are classified correctly.

\noindent\begin{center}
  \begin{tcolorbox}[colback=black!5, colframe=black!20, width=1.0\linewidth, arc=0.5mm, boxrule=0.8pt, left=1mm, right=1mm, top=0.5mm, bottom=0.5mm, boxsep=0.5mm]
    \textbf{Key Findings of RQ3:}
    \begin{itemize}
      \item Regarding the total number of classes, LMs' ability to identify salient classes does not deteriorate monotonically. Instead, their performance shows a U-shaped pattern, initially declining before recovering. 
      \item  As the number of salient classes per commit increases, the performance of both LMs and baselines drops significantly. Consequently, commits containing multiple salient classes remain the primary challenge for salient class identification.
      \item As diff token length increases, the ability of LMs to identify the complete set of salient classes in a commit declines.
  \end{itemize}
  \end{tcolorbox}
\end{center}

\subsection{Qualitative Analysis of LM Failure Cases (RQ4)}
This subsection presents a qualitative investigation into the failure cases of LMs when identifying salient classes. Unlike the aggregate metrics in the previous RQs, it focuses on the mechanisms behind errors. %We identify scenarios in which LMs still struggle when relying only on raw code diff and separate different decision mechanisms that lead predictions away from the gold salient class labels. 
For this analysis, we use GPT-5.4 zero-shot as a representative strong configuration and take commits whose prediction sets differ from gold salient class labels as candidate failure cases. Based on a 95\% confidence level and a 0.05 margin of error, we determine the sample size needed for manual analysis from the candidate failure cases in ApacheJavaCM and finally sample 328 instances for error attribution. For each sample, we inspect the commit message, code diff, distribution of modified lines, GPT-5.4 prediction, and ISC baseline output.

To ensure the objectivity and reliability of manual error attribution, we recruited three independent annotators with software engineering backgrounds and more than three years of programming experience. The annotators comprised two PhD students in software engineering and one industrial expert. The three participants first independently analyzed each sample and recorded the main error cause. They then discuss samples with disagreements, merge semantically similar explanations, and form the final classification. Rather than treating all errors as homogeneous failures, we summarize representative failure cases into the following three categories.

\subsubsection{Category I: Implicit dependency and missing structural context}

The first type of error comes from implicit dependencies and missing structured program context. In such commits, LMs are often drawn to downstream classes with larger code changes, while overlooking the root class that actually triggers the cascade of coupled modifications. Since our method only provides raw code diff to the LM and does not provide call graphs, inheritance relations, dependency edges, or reference paths between classes, the LM has difficulty judging whether a newly added or heavily rewritten class is the source of the change or a downstream modification caused by an upstream class. In contrast, the ISC baseline explicitly relies on static structural features and can therefore give correct results in some of such cases.

A typical case comes from Pinot commit \texttt{a020799a15cf}, whose commit message is ``[PINOT-3430] Moving ControllerLeaderLocator to core, since other realtime classes are there.'' The gold labels mark \codeid{ControllerLeaderLocator} as positive, while GPT-5.4 zero-shot predicts \codeid{ServerSegmentCompletionProtocolHandler}; the ISC baseline correctly predicts \codeid{ControllerLeaderLocator}. At the surface of the code diff, \codeid{ServerSegmentCompletionProtocolHandler} is a newly added class with 121 lines, which appears much more salient than \codeid{ControllerLeaderLocator}, which changes only 4 lines. However, the new class internally obtains the controller leader through \codeid{ControllerLeaderLocator.getInstance()}, and it is a downstream adaptation introduced after \codeid{ControllerLeaderLocator} is moved to the core module. In other words, GPT-5.4 misidentifies the downstream implementation class with the largest modification volume and a complete semantic unit as the salient class, while ISC can use structured dependency signals to identify the true root class. This case indicates that when movement across modules, dependency reconnection, or call-chain propagation exists, LMs that rely only on code diffs may still lack the necessary view of program structure.

\subsubsection{Category II: Attention diffusion under long diffs}

The second type of error occurs in long code diffs. Such commits often contain multiple files, multiple candidate classes, and many formatting adjustments, interface migrations, test synchronizations, or peripheral document changes. The class that truly carries the core change semantics may appear in only a few key lines, but these lines are diluted by many similar context fragments, making it difficult for the LM to stably anchor on the most important class in long inputs. This phenomenon is consistent with the result in RQ3 that increasing diff token length reduces LM \textit{CmtAccuracy}.

For example, Flink commit \texttt{ed4b94001fe6} has the commit message ``[FLINK-11451][table] Port *\allowbreak{}QueryConfig and TableDescriptor to flink-\allowbreak{}table-\allowbreak{}api-\allowbreak{}java.''

The gold salient class labels are \codeid{QueryConfig} and \codeid{TableDescriptor}, while GPT-5.4 zero-shot predicts \codeid{QueryConfig} and \codeid{StreamQueryConfig}, missing \codeid{TableDescriptor}. From the code diff structure, the commit involves 244 modified lines and multiple similar classes. \codeid{StreamQueryConfig} adds 86 lines, the old Scala \texttt{queryConfig.scala} removes 92 lines, whereas \codeid{TableDescriptor} contains only 9 key migration lines. The LM successfully identifies \codeid{QueryConfig}, but further allocates attention to \codeid{StreamQueryConfig}, which has a larger modification volume and more complete text, and fails to retain \codeid{TableDescriptor}, which is also explicitly mentioned in the commit message but has a shorter code diff fragment. This case shows that the modification scale and repeated context in long code diffs can interfere with the retention of a small number of key signals by the LM, leading to predictions that appear reasonable but are incomplete at the commit level.

\subsubsection{Category III: Omission of root causes under reactive implementation changes}

\textcolor{black}{A salient class may be the root-cause or primary-target class that motivates a commit}, but the visible code diff mainly appears in classes that are changed reactively to implement the fix. In these commits, LMs may follow the most explicit added interface, traversal condition, or control-flow logic and select the class where the repair is materialized. However, the salient class is the class whose behavior or role makes the change necessary. The LM therefore shifts from identifying the root cause to identifying repair carriers. This category differs from the first category because the wrongly predicted class is not merely a downstream dependency in a structural propagation chain; instead, it is a reactive implementation carrier introduced or adjusted to handle the root-cause class.

JMeter commit \texttt{0b69ec164ac5} illustrates this issue. Its message is ``Bug 60654 - Validation Feature : Be able to ignore BackendListener.'' The gold salient class label is \codeid{BackendListener}, while GPT-5.4 zero-shot predicts \codeid{Backend} and \codeid{TreeClonerForValidation}; the ISC baseline also predicts \codeid{Backend}. In the code diff, \codeid{BackendListener} changes only three lines by implementing the newly added \codeid{Backend} interface. Meanwhile, \codeid{Backend} is introduced as a marker interface, and \codeid{TreeClonerForValidation} adds the validation option and the conditional logic that skips objects implementing \codeid{Backend}. These two classes are modified so that the validation procedure can recognize and filter \codeid{BackendListener}. They therefore act as repair carriers for the underlying issue. The class that motivates the change remains \codeid{BackendListener}, because it is the component that should be ignored during validation and the class explicitly named in the issue. GPT-5.4 follows the visible interface and traversal logic, but it fails to trace these reactive implementation changes back to \codeid{BackendListener}. This case shows that LMs that rely only on code diffs may confuse classes that carry the repair with classes that constitute the modification root.

\noindent\begin{center}
  \begin{tcolorbox}[colback=black!5, colframe=black!20, width=1.0\linewidth, arc=0.5mm, boxrule=0.8pt, left=1mm, right=1mm, top=0.5mm, bottom=0.5mm, boxsep=0.5mm]
    \textbf{Key Findings of RQ4:}
    \begin{itemize}
      \item The failures of LMs on salient class identification do not arise from a single cause. They can be summarized into three categories: missing implicit dependencies make the LM prefer downstream classes with larger modification volume, long code diffs weaken stable LM attention to a small number of key lines, and reactive implementation changes make the LM confuse repair carriers with root-cause classes.
      \item The first two categories reveal real room for improving LMs that rely only on code diffs. Future methods can introduce call graphs, dependencies between classes, inheritance relations, or change propagation paths to supplement information about program structure missing from raw code diff.
      \item The third category shows that LMs need stronger tracing of root causes between the class that motivates the change and the classes that implement the corresponding repair. Future methods can add explicit constraints, ranking strategies, or post-processing rules that prefer root-cause classes when repair carriers and modification roots coexist in the same commit.
  \end{itemize}
  \end{tcolorbox}
\end{center}

\section{Threats to Validity}\label{sec:threats}
We discuss potential threats to validity from four perspectives: internal validity, external validity, construct validity, and conclusion validity, along with the mitigation measures adopted in our experimental design.

\subsection{Internal Validity}
Internal validity concerns whether the experimental process itself may introduce bias. 
%First, our label construction follows the salient class definition of ISC \cite{huang2022isc}, which treats modified classes explicitly mentioned in the commit message as salient classes. This rule ensures consistency with prior work in task setting, but it also makes the evaluation target different from a broad code-importance judgment over all modified classes. In particular, when a root-cause class and the repair carriers modified to handle it appear in the same commit, the experiment evaluates whether a method can recover the root class aligned with the commit message rather than whether it can identify every class that carries implementation logic. To mitigate ambiguity in applying this rule, we strictly apply unified class name matching and filtering rules during dataset construction.

First, the ISC baseline~\cite{huang2022isc} is not reproduced end-to-end from the complete replication package of the original paper. Instead, following the GBSCI protocol~\cite{ren2024gbsci} for reproducing ISC, we use the trained and directly executable ISC model released by Huang \textit{et al}. in another work~\cite{huang2022changepatterns} to generate results. This choice keeps our baseline consistent with the published reproduction protocol, but it may still differ in details from the original ISC experimental environment. To reduce this influence, all methods are evaluated on ApacheJavaCM, using the same input samples, labels, and metric calculation scripts.

Second, LM calls may be affected by decoding randomness, API version, and output format parsing. We set the temperature of all LMs to 0 and keep the same input format, task definition, and output constraints across LMs and prompting strategies. For LM outputs, we use deterministic post-processing scripts to extract predicted classes and retain raw responses for later inspection. Few-shot examples and prompt templates are fixed before the experiments and are not adjusted according to test results, avoiding additional bias toward specific LMs or samples.

\subsection{External Validity}
External validity concerns whether the conclusions can generalize to other projects, languages, and LMs. ApacheJavaCM is constructed from Java projects in the ApacheCM dataset. The results therefore first apply to open-source projects whose main language is Java, whose commit messages are relatively standardized, and whose class names can be recognized by CamelCase rules. For non-Java languages, commits involving multiple languages, private industrial projects, or repositories with different commit message styles, LM performance may change.

In addition, we select GPT-5.4, DeepSeek-V3.2, and Qwen3.5-9B to represent closed-source LLM, open-source LLM, and open-source SLM usage scenarios, respectively. Although these three LMs cover common deployment forms, they cannot represent all LMs. As LM architectures, training corpora, and alignment methods continue to evolve, other LMs may behave differently on salient class identification.

\subsection{Construct Validity}
Construct validity concerns whether the experimental design accurately measures the target theoretical concept. The main threat comes from how RQ3 operationalizes characteristics of commits. We use the total number of classes in a commit, the number of salient classes, and diff token length for grouped analysis. The first two are consistent with the experimental settings of ISC and GBSCI, whereas the latter characterizes the variation in input scale faced by LMs that directly process code diff. However, these three dimensions can only approximate observable characteristics of a commit and cannot fully represent its semantic difficulty. Factors such as modification type, call depth between classes, proportion of test code, degree of refactoring, and project historical context may also affect identification results.

\subsection{Conclusion Validity}
Conclusion validity concerns whether the experimental conclusions are sufficiently supported by data. In RQ1 to RQ3, we report not only a single metric but also multiple metrics and results under different groups, reducing misjudgment caused by accidental fluctuations in one metric. 

Part of the analysis still depends on the current LM versions. Closed-source LLMs may be updated on the server side. Even with the same prompts and decoding parameters, future repetitions may obtain slightly different outputs. To enhance inspectability, we have released the dataset, prompt templates, and evaluation scripts, enabling subsequent researchers to evaluate the results using the same inputs and post-processing logic~\cite{xiong2026replication}. For answering RQ4, error attribution still contains a degree of human judgment. Therefore, during case selection, we jointly inspect gold labels, model outputs, ISC outputs, commit messages, and code diffs to reduce the influence of any single evidence source on the conclusion.

\section{Conclusions and Future Work}\label{sec:conclusion}
This work presents the first empirical study of the performance of LMs in salient class identification. Unlike existing methods such as ISC and GBSCI, which depend on static analysis, feature extraction, or graph construction, our study provides raw code diffs to LMs through natural language prompts and asks them to determine whether each modified class is a salient class. We construct ApacheJavaCM from ApacheCM, yielding 7,911 commits and 25,914 labeled classes. Using ApacheJavaCM, we conduct a comprehensive experimental evaluation of LM performance on the salient class identification task and compare with the strongest reproducible baseline. The evaluation covers two LLMs and one SLM, three prompting strategies, eight evaluation metrics, and three dimensions of commit characteristics.

LMs outperform Most-Modified-Class and the ISC baseline without static analysis at the repository level, graph construction, or in-project training. The effects of LM selection and prompting strategy are not fully determined by the parameter scale. GPT-5.4 remains ahead on overall classification quality metrics such as \textit{MCC}, while the 9B-parameter open-source SLM Qwen3.5-9B achieves \textit{CmtAccuracy} comparable to GPT-5.4 under few-shot prompting. 
\textcolor{black}{This result supports locally deployable SLMs as feasible implementation options for this salient class identification task, particularly in settings where lower cost or stronger code privacy protection is desired.}
%This result suggests that locally deployable SLMs are practical for salient class identification, a relatively focused code understanding task, and may reduce the cost of closed-source LLMs and code privacy risks.
Grouped experiments further show that the total number of classes in a commit does not cause monotonic performance decline, commits with multiple salient classes remain a common difficult scenario for all methods, and increasing diff token length weakens the identification performance of LMs.

There are several directions that remain for future work. Structured program context such as call graphs, class dependency relations, inheritance relations, and historical co-change information can be introduced into LM input to mitigate insufficient implicit dependency reasoning when relying only on code diff. For long code-diff scenarios, future studies could explore input compression strategies based on retrieval, slicing, chunk summarization, or key-hunk localization, allowing LMs to prioritize code fragments that truly capture the intent of the change. Further experiments across a broader range of programming languages, industrial projects, and LM versions are needed, along with studies on deployment, confidence calibration, and human--LM collaboration workflows for integrating local SLMs into code review toolchains.

\section*{Data Availability}
The replication package of this study has been made available at~\cite{xiong2026replication}.

\section*{Acknowledgments}
This work has been partially supported by the National Natural Science Foundation of China (NSFC) under Grant Nos. 92582203 and 62032016.

\bibliographystyle{ieeetr}
\bibliography{ref}

\end{sloppypar}
\end{document}